\documentclass[twocolumn,showpacs,preprintnumbers,pra,hyperref]{revtex4}
\usepackage{amssymb}
\usepackage{amsfonts}
\usepackage{amsmath}
\usepackage{graphicx}

\begin{document}

\title{Transmission Character of General Function Photonic Crystals}
\author{Xiang-Yao Wu$^{a}$\thanks{%
E-mail: wuxy2066@163.com}, Bo-Jun Zhang$^{a}$, Jing-Hai
Yang$^{a}$, Si-Qi Zhang$^{a}$\\ Xiao-Jing Liu$^{a}$, Jing
Wang$^{a}$, Nuo Ba$^{a}$ and Zhong Hua$^{a}$}
\affiliation{$^{a}${\small Institute of Physics, Jilin Normal
University, Siping 136000, China}}

\begin{abstract}
In the paper, we present a new general function photonic crystals
(GFPCs), which refractive index of medium is a arbitrary function
of space position. Unlike conventional photonic crystals (PCs),
which structure grow from two mediums $A$ and $B$, with different
constant refractive indexes  $n_{a}$ and $n_{b}$. Based on Fermat
principle, we give the motion equations of light in
one-dimensional GFPCs, and calculate its transfer matrix, which is
different from the conventional PCs. We choose the linearity
refractive index function for two mediums $A$ and $B$, and find
the transmissivity of one-dimensional GFPCs can be much larger or
smaller than $1$ for different slope linearity refractive index
function, which is different from the transmissivity of
conventional PCs (its transmissivity is in the range of $0$ and
$1$). Otherwise, we study the effect of different incident angles,
the number of periods and optical thickness on the transmissivity,
and obtain some new results different from the conventional PCs.
\end{abstract}

\maketitle

\maketitle {\bf 1. Introduction} \vskip 8pt

Photonic crystals (PCs) are composite structures with a periodic
arrangement of materials with different refractive indices in
one-dimension (1D), two-dimension (2D) or three-dimension (3D).Due
to the introduced periodicity, multiple Bragg scatterings from
each unit cell may open a photonic band gaps (PBGs), analogous to
the electronic band gaps in semiconductors, with in which the
propagation of electromagnetic (EM) waves is completely forbidden.
The existence of PBGs will lead to many interesting phenomena,
e.g., modification of spontaneous emission [1-5] and photon
localization [6-10]. Thus numerous applications of photonic
crystals have been proposed in improving the performance of
optoelectronic and microwave devices such as high-efficiency
semiconductor lasers, right emitting diodes, wave guides, optical
filters, high-Q resonators, antennas, frequency-selective surface,
optical limiters and amplifiers [11-14]. These applications would
be significantly enhanced if the band structure of the photonic
crystal could be tuned.

For the conventional PCs, the photonic band gaps remain fixed once
the PCs have been fabricated. If the band gaps of the photonic
crystals could be tuned the applications would be significantly
enhanced. A practical scheme for tuning the band gap was proposed
by Busch and John [15]. It has been demonstrated theoretically and
experimentally that PCs with liquid crystal infiltration exhibit
tunability on applying an external electric field [15]or changing
the temperature [16-20]. If the constituent materials of PCs have
magnetic permeabilities dependent on the external magnetic field,
the photonic band gaps [PBGs] can be altered by changing the
external magnetic field [21-28]. An electric or magnetic field
changes the PBGs easily than the temperature.

In Ref. [29], we have proposed special function photonic crystals,
which the medium refractive index is the function of space
position, but the function value of refractive index is equal at
two endpoints of every medium $A$ and $B$, and obtain some new
results different from the the conventional PCs. In this paper, we
present a new general function photonic crystals (GFPCs), which
refractive index is a arbitrary function of space position
(needless refractive index same at two endpoint). Unlike
conventional photonic crystals (PCs), which structure grow from
two materials, A and B, with different dielectric constants
$\varepsilon_{A}$ and $\varepsilon_{B}$. Firstly, we give the
motion equation of light in one-dimensional GFPCs according to
Fermat principle. Secondly, we calculate the transfer matrix for
the one-dimensional GFPCs, which is different from the transfer
matrix of the conventional PCs. Finally, we give the dispersion
relation, band gap structure and transmissivity. We choose the
linearity refractive index function for two  medium $A$ and $B$,
and find the transmissivity of GFPCs can be much larger or smaller
than $1$ for the different slope linearity refractive index
function, which is different from the transmissivity of
conventional PCs (its transmissivity is in the range of $0$ and
$1$). Otherwise, we study the effect of different incident angles,
the number of periods and optical thickness on the transmissivity,
and obtain some new results. By the calculation, we find the
conventional PCs is the special case of the GFPCs.

\vskip 8pt

{\bf 2. The light motion equation in general function photonic
crystals} \vskip 8pt

For the general function photonic crystals, the medium refractive
index is a periodic function of the space position, which can be
written as $n(z)$, $n(x, z)$ and $n(x, y, z)$ corresponding to
one-dimensional, two-dimensional and three-dimensional function
photonic crystals. In the following, we shall deduce the light
motion equations of the one-dimensional general function photonic
crystals, i.e., the refractive index function is $n=n(z)$,
meanwhile motion path is on $xz$ plane. The incident light wave
strikes plane interface point $A$, the curves $AB$ and $BC$ are
the path of incident and reflected light respectively, and they
are shown in FIG. 1.
\begin{figure}[tbp]
\includegraphics[width=8.5 cm]{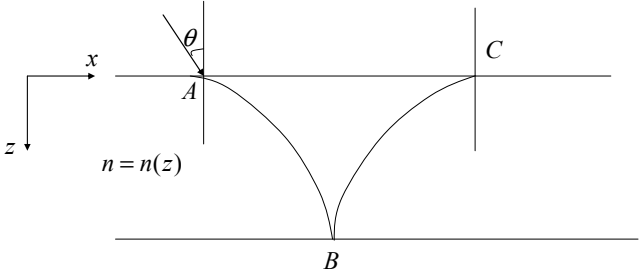}
\caption{The motion path of light in the medium of refractive
index $n(z)$.}
\end{figure}

The light motion equation can be obtained by Fermat principle, it
is
\begin{eqnarray}
\delta\int^{B}_{A}n(z) ds=0.\label{1}
\end{eqnarray}
In the two-dimensional transmission space, the line element $ds$
is
\begin{eqnarray}
ds=\sqrt{(dx)^{2}+(dz)^{2}}=\sqrt{1+\dot{z}^{2}}dx,
\end{eqnarray}
where $\dot{z}=\frac{dz}{dx}$, then Eq. (1) becomes
\begin{eqnarray}
\delta\int^{B}_{A}n(z)\sqrt{1+(\dot{z})^{2}}dx=0.
\end{eqnarray}
The Eq. (3) change into
\begin{eqnarray}
\int^{B}_{A}(\frac{\partial(n(z)\sqrt{1+\dot{z}^{2}})}{\partial
z}\delta z+\frac{\partial(n(z)\sqrt{1+\dot{z}^{2}})}{\partial
\dot{z}}\delta\dot{z})dx=0,
\end{eqnarray}
At the two end points $A$ and $B$, their variation is zero, i.e.,
$\delta z (A)=\delta z (B)=0$. For arbitrary variation $\delta z$, the Eq. (4) becomes \\
\begin{eqnarray}
&&\frac{dn(z)}{dz}\sqrt{1+\dot{z}^{2}} -\frac{d n(z)}{d
z}\dot{z}^{2}(1+\dot{z}^{2})^{-\frac{1}{2}}
\nonumber\\&&-n(z)\frac{\ddot{z}\sqrt{1+\dot{z}^{2}}
-\dot{z}^{2}\ddot{z}(1+\dot{z}^{2})^{-\frac{1}{2}}}{1+\dot{z}^{2}}
=0,
\end{eqnarray}
simplify Eq. (5), we have
\begin{eqnarray}
\frac{d n(z)}{n(z)} = \frac{\dot{z}d\dot{z}}{1+\dot{z}^{2}}.
\end{eqnarray}\\
The Eq. (6) is light motion equation in one-dimensional function
photonic crystals.
\begin{figure}[tbp]
\includegraphics[width=8.5 cm]{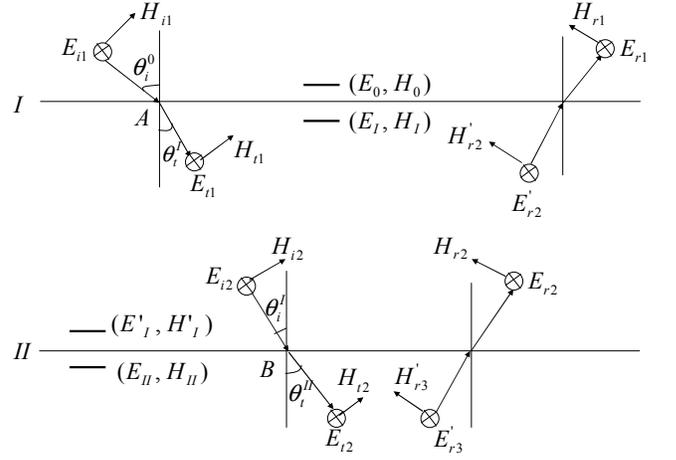}
\caption{The light transmission and electric magnetic field
distribution figure in FIG.1 medium.} \label{Fig1}
\end{figure}

{\bf 3. The transfer matrix of one-dimensional general function
photonic crystals}

In this section, we should calculate the transfer matrix of
one-dimensional general function photonic crystals. In fact, there
is the reflection and refraction of light at a plane surface of
two media with different dielectric properties. The dynamic
properties of the electric field and magnetic field are contained
in the boundary conditions: normal components of $D$ and $B$ are
continuous; tangential components of $E$ and $H$ are continuous.
We consider the electric field perpendicular to the plane of
incidence, and the coordinate system and symbols as shown in FIG.
2.

On the two sides of interface I, the tangential components of
electric field $E$ and magnetic field $H$ are continuous, there
are

\begin{eqnarray}
\left \{ \begin{array}{ll}
 E_{0}=E_{I}=E_{t1}+E'_{r2}\\
H_{0}=H_{I}=H_{t1}\cos\theta_{t}^{I}-H'_{r2}\cos\theta_{t}^{I}.
\end{array}
\right.
\end{eqnarray}
On the two sides of interface II, the tangential components of
electric field $E$ and magnetic field $H$ are continuous, and give
\begin{eqnarray}
\left \{ \begin{array}{ll}
 E_{II}=E'_{I}=E_{i2}+E_{r2}\\
H_{II}=H'_{I}=H_{i2}\cos\theta_{i}^{I}-H_{r2}\cos\theta_{i}^{I},
\end{array}
\right.
\end{eqnarray}
the electric field ${E_{t1}}$ is
\begin{eqnarray}
E_{t1}=E_{t10}{e^{i(k_{x}x_{A}+k_{z}z)}|_{z=0}}=E_{t10}e^{i\frac{\omega}{c}n(0)\sin\theta_{t}^{I}x_{A}},
\end{eqnarray}
and the electric field ${E_{i2}}$ is
\begin{eqnarray}
E_{i2}&=&E_{t10}{e^{i(k'_{x}x_{B}+k'_{z} z)}|_{z=b}}
\nonumber\\&=&E_{t10}e^{i\frac{\omega}{c}n(b)(\sin\theta_{i}^{I}x_{B}+\cos\theta_{i}^{I}
b)}.
\end{eqnarray}
Where $x_{A}$ and $x_{B}$ are $x$ component coordinates
corresponding to point $A$ and point $B$. We should give the
relation between $E_{i2}$ and $E_{t1}$. By integrating the two
sides of Eq. (6), we can obtain the coordinate component $x_{B}$
of point $B$
\begin{eqnarray}
\int^{n(z)}_{n(0)}\frac{dn(z)}{n(z)}=\int^{k_{z}}_{k_{0}}\frac{\dot{z}d\dot{z}}{1+\dot{z}^{2}},
\end{eqnarray}
to get
\begin{eqnarray}
k_z^2=(1+k_0^2)(\frac{n(z)}{n(0)})^2-1,
\end{eqnarray}
and
\begin{eqnarray}
dx=\frac{dz}{\sqrt{(1+k_{0}^{2})(\frac{n(z)}{n(0)})^{2}-1}}.
\end{eqnarray}
where $k_{0}=\cot\theta_{t}^{I}$ and $k_{z}=\frac{d z}{d x}$ From
Eq. (12), there is $n(z)>n(0)\sin\theta^{I}_{t}$. and the
coordinate $x_{B}$ is
\begin{eqnarray}
x_{B}=x_{A}+\int^{b}_{0}\frac{dz}{\sqrt{(1+k_{0}^{2})(\frac{n(z)}{n(0)})^{2}-1}},
\end{eqnarray}
where $b$ is the medium thickness of FIG. 1 and FIG. 2.\\
By substituting Eqs. (9) and (14)into (10), and using the equality
\begin{eqnarray}
 n(0)\sin\theta_{t}^{I}=n(b)\sin\theta_{i}^{I},
\end{eqnarray}
  we have
\begin{eqnarray}
E_{i2}&=&E_{t1}e^{i{\delta}_{b}},
\end{eqnarray}
where
\begin{figure}[tbp]
\includegraphics[width=8 cm]{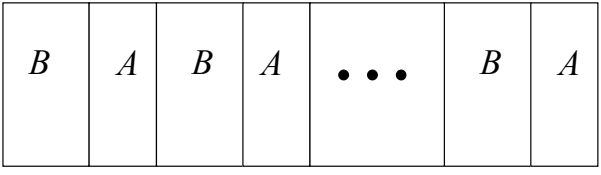}
\caption{The structure $(BA)^{N}$ of the general function photonic
crystals.}
\end{figure}
\begin{eqnarray}
\delta_{b}=\frac{\omega}{c}n_{b}(b)(\cos\theta_{i}^{I}b+\sin\theta_{i}^{I}
\int^{b}_{0}\frac{dz}{\sqrt{\frac{n_{b}^{2}(z)}{n_{0}^{2}\sin^{2}\theta_{i}^{0}}-1}}),
\end{eqnarray}
and similarly
\begin{eqnarray}
E'_{r2}=E_{r2}e^{i\delta_{b}}.
\end{eqnarray}
Substituting Eqs. (16) and (18) into (7) and (8), and using
$H=\sqrt{\frac{\varepsilon_{0}}{\mu_{0}}}nE$, we obtain
\begin{eqnarray}
\left(%
\begin{array}{c}
  E_{I} \\
  H_{I} \\
\end{array}%
\right)&=&M_{B}\left(%
\begin{array}{c}
  E_{II} \\
  H_{II} \\
\end{array}%
\right),
\end{eqnarray}
where
\begin{eqnarray}
M_{B}=\left(%
\begin{array}{cc}
 \cos\delta_{b} & -\frac{i\sin\delta_{b}}{\sqrt{\frac{\varepsilon_{0}}{\mu_{0}}}n_{b}(b)\cos\theta_{i}^{I}} \\
 -in_{b}(0)\sqrt{\frac{\varepsilon_{0}}{\mu_{o}}}\cos\theta_{t}^{I}\sin\delta_{b}
 & \frac{n_{b}(0)\cos\theta_{t}^{I}\cos\delta_{b}}{n_{b}(b)\cos\theta_{i}^{I}}\\
\end{array}%
\right),
\end{eqnarray}
The Eq. (20) is the transfer matrix $M$ in the medium of FIG. 1
and FIG. 2. By refraction law, we can obtain
\begin{eqnarray}
\sin\theta^{I}_{t}=\frac{n_{0}}{n(0)}\sin\theta^{0}_{i},\cos\theta^{I}_{t}
=\sqrt{1-\frac{n_{0}^{2}}{n^{2}(0)}\sin^{2}\theta^{I}_{t}},
\end{eqnarray}
where $n_0$ is air refractive index, and $n(0)=n(z)|_{z=0}$. Using
Eqs. (15) and (21), we can calculate $\cos\theta_{i}^{I}$.

\begin{figure}[tbp]
\includegraphics[width=8 cm]{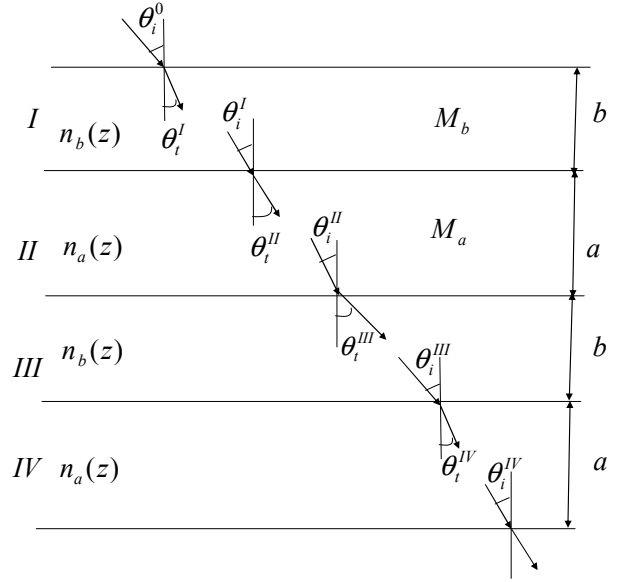}
\caption{The two periods transmission figure of light in general
function photonic crystals.}
\end{figure}
{\bf 4. The structure of one-dimensional general function photonic
crystals} \vskip 8pt

In section 3, we obtain the $M$ matrix of the half period. We know
that the conventional photonic crystals is constituted by two
different refractive index medium, and the refractive indexes are
not continuous on the interface of the two mediums. We could
devise the one-dimensional general function photonic crystals
structure as follows: in the first half period, the refractive
index distributing function of medium $B$ is $n_{b}(z)$. and in
the second half period, the refractive index distributing function
of medium $A$ is $n_{a}(z)$, corresponding thicknesses are $b$ and
$a$, respectively. Their refractive indexes satisfy condition
$n_{b}(b)\neq n_{a}(0)$, their structure are shown in FIG. 3, and
FIG. 4. The Eq. (20) is the half period transfer matrix of medium
$B$. Obviously, the half period transfer matrix of medium A is
\begin{eqnarray}
M_{A}=\left(%
\begin{array}{cc}
 \cos\delta_{a} & -\frac{i\sin\delta_{a}}{\sqrt{\frac{\varepsilon_{0}}{\mu_{0}}}n_{a}(a)\cos\theta_{i}^{II}} \\
 -in_{a}(0)\sqrt{\frac{\varepsilon_{0}}{\mu_{o}}}\cos\theta_{t}^{II}\sin\delta_{a}
 & \frac{n_{a}(0)\cos\theta_{t}^{II}\cos\delta_{a}}{n_{a}(a)\cos\theta_{i}^{II}}\\
\end{array}%
\right),
\end{eqnarray}
where
\begin{eqnarray}
\delta_{a}&=&\frac{\omega}{c}n_{a}(a)[\cos\theta^{II}_{i}\cdot a
\nonumber\\&&
+\sin\theta^{II}_{i}\int^{a}_{0}\frac{dz}{\sqrt{\frac{n_{a}^{2}(z)}{n_{0}^{2}\sin^{2}\theta_{i}^{0}}-1}}],
\end{eqnarray}
\begin{eqnarray}
\cos\theta^{II}_{t}
=\sqrt{1-\frac{n_{0}^{2}}{n_{a}^{2}(0)}\sin^{2}\theta_{i}^{0}},
\end{eqnarray}
and
\begin{eqnarray}
\sin\theta^{II}_{i}=\frac{n_{0}}{n_{a}(a)}\sin\theta_{i}^{0},
\end{eqnarray}
\begin{eqnarray}
\cos\theta^{II}_{i}
=\sqrt{1-\frac{n_{0}^{2}}{n_{a}^{2}(a)}\sin^{2}\theta_{i}^{0}}.
\end{eqnarray}
In one period, the transfer matrix $M$ is
\begin{eqnarray}
&&M=M_{B}\cdot M_{A}\nonumber\\
&&=\left(%
\begin{array}{cc}
  \cos\delta_{b} & \frac{-i\sin\delta_{b}}{\sqrt{\frac{\varepsilon_{0}}{\mu_{0}}}n_{b}(b)\cos\theta_{i}^{I}} \\
 -in_{b}(0)\sqrt{\frac{\varepsilon_{0}}{\mu_{o}}}\cos\theta_{t}^{I}\sin\delta_{b}
 & \frac{n_{b}(0)\cos\theta_{t}^{I}\cos\delta_{b}}{n_{b}(b)\cos\theta_{i}^{I}}\\
\end{array}%
\right) \nonumber\\&&
\left(%
\begin{array}{cc}
   \cos\delta_{a} & \frac{-i\sin\delta_{a}}{\sqrt{\frac{\varepsilon_{0}}{\mu_{0}}}n_{a}(a)\cos\theta_{i}^{II}} \\
 -in_{a}(0)\sqrt{\frac{\varepsilon_{0}}{\mu_{o}}}\cos\theta_{t}^{II}\sin\delta_{a}
 & \frac{n_{a}(0)\cos\theta_{t}^{II}\cos\delta_{a}}{n_{a}(a)\cos\theta_{i}^{II}}\\
\end{array}%
\right).
\end{eqnarray}
The form of the GFPCs transfer matrix $M$ is more complex than the
conventional PCs. The angle $\theta_{t}^{I}$, $\theta_{i}^{I}$,
$\theta_{t}^{II}$ and $\theta_{i}^{II}$ are shown in Fig. 4. The
characteristic equation of GFPCs is
\begin{eqnarray}
\left(%
\begin{array}{c}
  E_{1} \\
  H_{1} \\
\end{array}%
\right)&=&M_{1}M_{2}\cdot\cdot\cdot M_{N}
\left(%
\begin{array}{c}
  E_{N+1} \\
  H_{N+1} \\
\end{array}%
\right) \nonumber\\&=&M_{b}M_{a}M_{b}M_{a}\cdot\cdot\cdot M_{b}M_{a}\left(%
\begin{array}{c}
  E_{N+1} \\
  H_{N+1} \\
\end{array}%
\right)
\nonumber\\&=&M\left(%
\begin{array}{c}
  E_{N+1} \\
  H_{N+1} \\
\end{array}%
\right)=\left(%
\begin{array}{c c}
  A &  B \\
 C &  D \\
\end{array}%
\right)
 \left(%
\begin{array}{c}
  E_{N+1} \\
  H_{N+1} \\
\end{array}%
\right).
\end{eqnarray}
Where $N$ is the number of period.

\begin{figure}[tbp]
\includegraphics[width=9 cm]{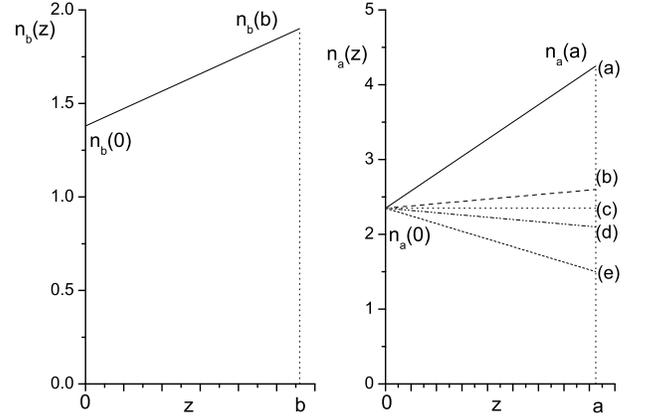}
\caption{The line refractive index function in a period. The
straight line (a-e) are corresponding to different line functions
for medium $A$.}
\end{figure}

{\bf 5. The dispersion relation, band gap structure and
transmissivity} \vskip 8pt

With the transfer matrix $M$ (Eq. (27)), we can study the
dispersion relation and band gap structure of the function
photonic crystals.\\
Since
\begin{eqnarray}
\left(%
\begin{array}{c}
E_{N} \\
H_{N} \\
\end{array}%
\right)=M\left(%
\begin{array}{c}
  E_{N+1} \\
  H_{N+1} \\
\end{array}%
\right)=M_{b}M_{a}\left(%
\begin{array}{c}
  E_{N+1} \\
  H_{N+1} \\
\end{array}%
\right).
\end{eqnarray}
By Bloch law, there is
\begin{eqnarray}
\left(%
\begin{array}{c}
  E_{N} \\
 H_{N}\\
\end{array}%
\right)=e^{-ikd}\left(%
\begin{array}{c}
  E_{N+1} \\
  H_{N+1} \\
\end{array}%
\right),
\end{eqnarray}
where $d=b+a$. With Eqs. (29) and (30), there is
\begin{eqnarray}
\left(%
\begin{array}{c}
  E_{N} \\
  H_{N} \\
\end{array}%
\right)=M_{b}M_{a}\left(%
\begin{array}{c}
  E_{N+1} \\
  H_{N+1} \\
\end{array}%
\right)=e^{-ikd}\left(%
\begin{array}{c}
  E_{N+1} \\
  H_{N+1} \\
\end{array}%
\right),
\end{eqnarray}
The non-zero solution condition of Eq. (31) is
\begin{eqnarray}
det(M_{b}M_{a}-e^{-ikd})=0,
\end{eqnarray}
i.e.,
\begin{widetext}
\begin{eqnarray}
\nonumber\\&&(\cos\delta_{b}\cos\delta_{a}-\frac{\eta_{a}}{\eta_{b}}\sin\delta_{b}\sin\delta_{a}-e^{-ikd})
(\cos\delta_{b}\cos\delta_{a}-\frac{\eta_{b}}{\eta_{a}}\sin\delta_{b}\sin\delta_{a}-e^{-ikd})
\nonumber\\&&+(-\frac{i}{\eta_{a}}\cos\delta_{b}\sin\delta_{a}-\frac{i}{\eta_{b}}\sin\delta_{b}\cos\delta_{a})
(-i\eta_{b}\sin\delta_{b}\cos\delta_{a}-i\eta_{a}\cos\delta_{b}\sin\delta_{a})=0.
\end{eqnarray}
\end{widetext}
Simplifying Eq.(33), we obtain the dispersion relation
\begin{eqnarray}
\cos
kd=\cos\delta_{b}\cos\delta_{a}-\frac{1}{2}(\frac{\eta_{b}}{\eta_{a}}+
\frac{\eta_{a}}{\eta_{b}})\sin\delta_{b}\sin\delta_{a}.
\end{eqnarray}
From Eq. (34), we can study the photonic dispersion relation and
band gap structure, and can obtain the transmission coefficient
$t$ from Eq. (28)
\begin{eqnarray}
t=\frac{E_{tN+1}}{E_{i1}}=\frac{2\eta_{0}}{A\eta_{0}+B\eta_{0}\eta_{N+1}+C+D\eta_{N+1}},
\end{eqnarray}
and transmissivity $T$ is
\begin{eqnarray}
T=t\cdot t^{*}.
\end{eqnarray}

\vskip 5pt {\bf 6. Numerical result} \vskip 5pt

In this section, we report  our numerical results of
transmissivity. We consider refractive indexes of the linearity
functions in a period, it is

\begin{eqnarray}
n_{b}(z)=n_{b}(0)+\frac{n_{b}(b)-n_{b}(0)}{b}z, \hspace{0.1in} 0
\leq z\leq b,
\end{eqnarray}
\begin{eqnarray}
n_{a}(z)=n_{a}(0)+\frac{n_{a}(a)-n_{a}(0)}{a}z, \hspace{0.1in} 0
\leq z\leq a,
\end{eqnarray}
Eqs. (37) and (38) are the refractive indexes distribution
function of two half period mediums $B$ and $A$ , which are shown
in FIG. 5. When the endpoint values $n_{b}(0)$, $n_{b}(b)$,
$n_{a}(0)$ and $n_{a}(a)$ are all given, the line refractive index
functions $n_{b}(z)$ and $n_{a}(z)$ are ascertained.

The main parameters are: the half period thickness $b$ and $a$,
the starting point refractive indexes $n_{b}(0)$ and $n_{a}(0)$,
and end point refractive indexes $n_{b}(b)$ and $n_{a}(a)$, the
optical thickness of the two mediums are equal, i.e.,
$n_{b}(0)b=n_{a}(0)a$.

\begin{figure}[tbp]
\includegraphics[width=9cm]{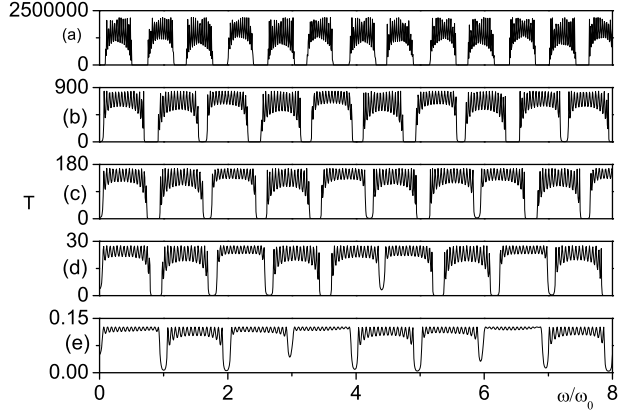}
\caption{Comparing the transmissivity of the GFPCs with different
line refractive index function: Fig. 6 (a-e) corresponding to Fig.
5 (a-e).}
\end{figure}

\begin{figure}[tbp]
\includegraphics[width=9cm]{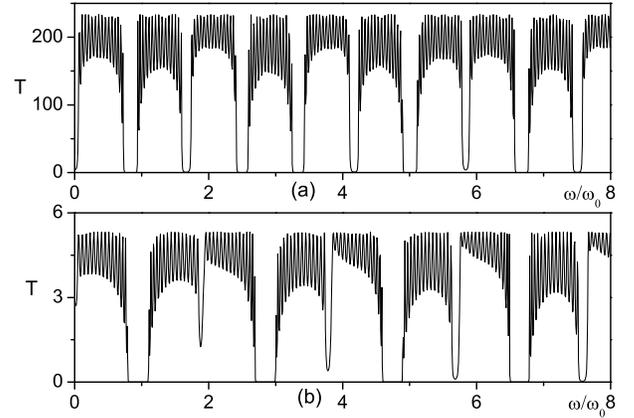}
\caption{Comparing the transmissivity of the GFPCs with different
line refractive index function for medium $B$. }
\end{figure}

In Fig.6, we take $n_{b}(0)=1.38$, $n_{b}(b)=1.9$, i.e., the
refractive index distribution of medium $B$ is confirmable,
$n_{a}(0)=2.35$, the incident angle $\theta_{i}^{0}=0$, the center
frequency $\omega_{0}=1.215\times10^{15} Hz$,
$\lambda_{0}=\frac{2\pi c}{\omega_{0}}$, the thickness $b=280 nm$
and $a=165 nm$ and the period number $N=16$. From Fig. 6(a-e), the
end point value of $n_{a}(a)$ are taken as: $4.25$, $2.6$, $2.35$,
$2.1$ and $1.5$ respectively. And the corresponding refractive
indexes distribution are straight lines (a), (b), (c), (d) and (e)
in Fig. 5. When the refractive index distribution of medium $B$
maintain unchanging, with the end point value $n_{a}(a)$ of medium
$A$ decrease, we can obtain the following results for the
one-dimensional GFPCs: (1) The maximum of transmissivity decrease.
From Fig.6 (a) to (e), their transmissivity maximums are
$2500000$, $900$, $180$, $30$ and $0.15$, respectively. (2) The
width of band gaps become narrow and the number of band gaps
decrease. (3)The transmissivity of one-dimensional GFPCs can be
much larger or smaller than $1$ for different slope linearity
refractive index function, which is different from the
transmissivity of conventional PCs (its transmissivity $T\leq 1$).
When we take $n_{b}{0}=n_{b}(b)$ and $n_{a}{0}=n_{a}(a)$, and
calculate the transmissivity by the GFPCs equations, i.e., Eqs.
(27) and (35), the calculate results of transmissivity is the same
as the conventional PCs. So, the conventional PCs is the special
case of the GFPCs.

In Fig. 7, we take $n_{a}(0)=2.35$, $n_{a}(a)=2.4$, i.e., the
refractive index distribution of medium $A$ is confirmable,
$n_{b}(0)=1.38$, and the end point value $n_{b}(b)$ of medium $B$
are taken as: $1.9$ and $1.5$. By the two refractive indexes
distributions of medium $B$, we obtain the transmissivity are
corresponding to Fig. 7(a) and (b). It can be found that when the
refractive index of medium $A$ keep unchanging, with the end point
value $n_{b}(b)$ of medium $B$ decrease, the transmissivity
maximum and the number of band gaps decrease, and the width of
band gaps increase.

In Fig. 8, Fig. 9 and Fig. 10, the main parameters are:
$n_{b}(0)=1.38$, $n_{b}(b)=1.9$, $n_{a}(0)=2.35$, $n_{a}(a)=2.6$.
In the inset of Fig. 8, we discuss the influence of period number
$N$ on the transmissivity. The results shown in Fig. 8 (a-c) are
calculated by the period number $N=12$, $N=16$ and $N=20$,
respectively. We can see that when the period number of the FGPCs
is small, for example, $N=12$ the maximum of transmission
intensity achieve $160$ [Fig. 8(a)]. When the period number
increase up to $20$ [Fig. 8(c)], the transmission intensity is
nearly $5000$, it is much higher than $N=12$, i.e., as the period
number increase, the transmission intensity increase.

Figure 9 shows that the effect of different incident angles on
transmission intensity. In Fig. 9 (a-d), the incident angles are:
$\theta_{i}^{0}=0$, $\theta_{i}^{0}=\frac{\pi}{6}$,
$\theta_{i}^{0}=\frac{\pi}{4}$ and $\theta_{i}^{0}=\frac{\pi}{3}$,
respectively. As we can see in Fig. 9 (a-d), when the incident
angle increase, the transmission intensity increase, and the
number of band gaps increase.

Figure 10 shows the variation of transmissivity due to the change
of the optical thickness. The optical thickness of $A$ and $B$
medium are equal, i.e., $n_{b}(0)b=n_{a}(0)a$. The results shown
in Fig. 10 (a-c) are calculated by the optical thickness as:
$\frac{\lambda_{0}}{8}$, $\frac{\lambda_{0}}{4}$ and
$\frac{\lambda_{0}}{2}$ respectively. It is found when the optical
thickness increase, the number of band gaps increase, while the
maximum of transmission intensity keep unchanged.

In the following, we consider two groups line refractive indexes
functions in a period, which are all decreasing linearly, they are
shown in Fig. 11. The first group main parameters are:
$n_{b}(0)=1.9$, $n_{b}(b)=1.8$, $n_{a}(0)=2.6$ and $n_{a}(a)=2.5$,
it is the solid line (a). The second group main parameters are:
$n_{b}(0)=1.5$, $n_{b}(b)=1.4$, $n_{a}(0)=2.4$ and $n_{a}(a)=2.3$,
it is the dash-dot line (b). The optical thickness are same:
$n_{b}(0)b=n_{a}(0)a=\frac{\lambda_{0}}{4}$.

Figure 12 is the transmission spectra. The Fig. 12 (a) correspond
to the first group line refractive indexes functions, the Fig. 12
(b) correspond to the second group line refractive indexes
function. Here, we notice that when the refractive indexes
function of the two mediums $A$ and $B$ are all decreasing, the
transmission intensity decrease and the width of band gaps become
wider.

In Fig. 13-15, the main parameters are: $n_{b}(0)=1.9$,
$n_{b}(b)=1.8$, $n_{a}(0)=2.6$ and $n_{a}(a)=2.5$.

In Fig. 13, we study the effect of period number $N$ on the
transmission intensity. The incident angle $\theta_{i}^{0}=0$, the
optical thickness $n_{b}(0)b=n_{a}(0)a=\frac{\lambda_{0}}{4}$.
They are shown in Fig. 13 (a-c), and the period number are $N=12$,
$N=16$ and $N=20$, respectively. It can be found that as period
number $N$ increases from $12$ to $20$, the transmission intensity
decrease.

In addition, we study the transmission intensity with different
incident angles as shown in Fig. 14. We take the period number
$N=16$, and optical thickness
$n_{b}(0)b=n_{a}(0)a=\frac{\lambda_{0}}{4}$. we can find when the
incident angle $\theta_{i}^{0}$ increase, the transmission
intensity decrease. For example, when the incident angle
$\theta_{i}^{0}=0$, the maximum of transmission intensity nearly
$0.08$ (Fig. 14(a)), and $\theta_{i}^{0}=\frac{\pi}{3}$, the
maximum of transmission intensity nearly $0.08$ (Fig. 14(d)).

In what follows, we will take into account the transmission at
different optical thickness, which is shown in Fig. 15. We also
choose $\theta_{i}^{0}=0$, $N=16$. When the optical thickness are
taken $\frac{\lambda_{0}}{8}$, $\frac{\lambda_{0}}{4}$ and
$\frac{\lambda_{0}}{2}$, the transmission spectra are
corresponding to Fig. 15 (a-c). we can find when the optical
thickness increases, the number of band gaps increase, the width
of band gaps become narrow, while the maximum of transmission
intensity keep unchanged.

\begin{figure}[tbp]
\includegraphics[width=9 cm]{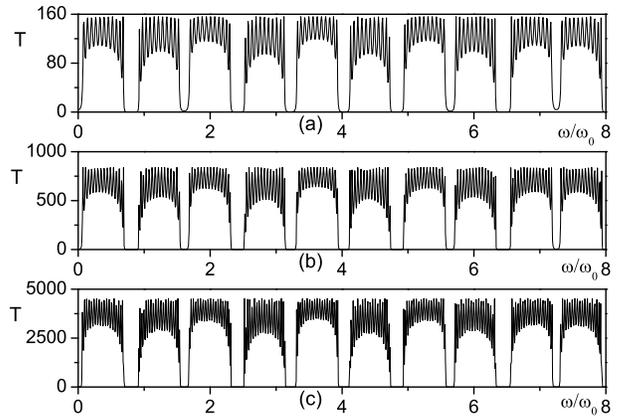}
\caption{Comparing the transmissivity of the function PCs with
different period number $N$: (a) N=12, (b) N=16 and (c) N=20.}

\end{figure}

\begin{figure}[tbp]
\includegraphics[width=8.5cm]{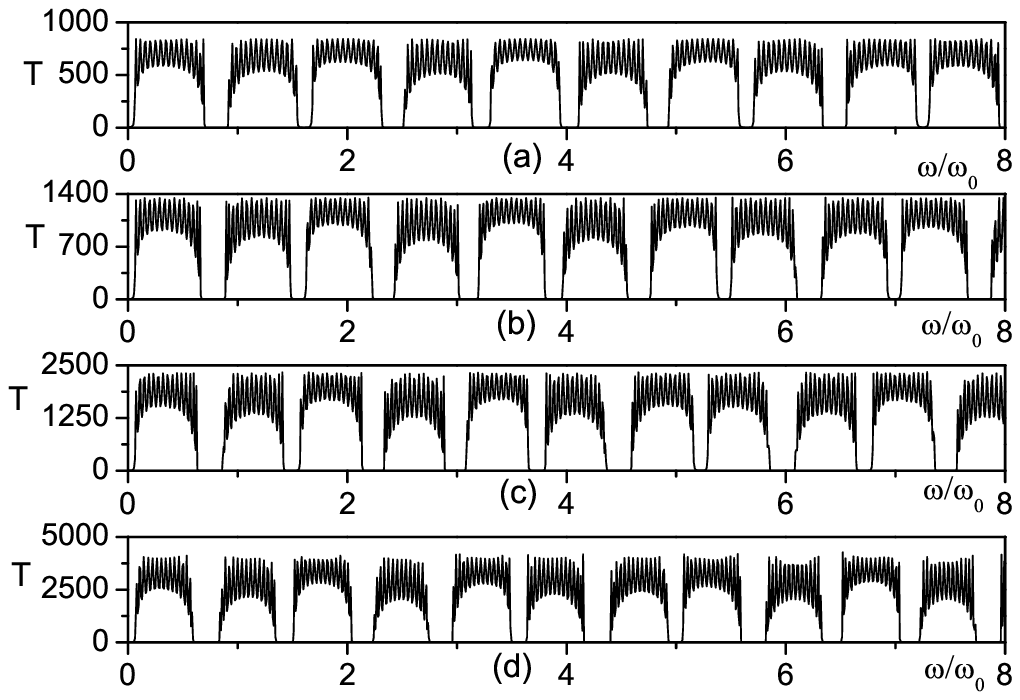}
\caption{Comparing the transmissivity of the GFPCs with different
incidence angle: (a)$\theta_{i}^{0}=0$, (b)
$\theta_{i}^{0}=\frac{\pi}{6}$,  (c)$\theta_{i}^{0}=\frac{\pi}{4}$
and (d) $\theta_{i}^{0}=\frac{\pi}{3}$.}
\end{figure}

\begin{figure}[tbp]
\includegraphics[width=8.5 cm]{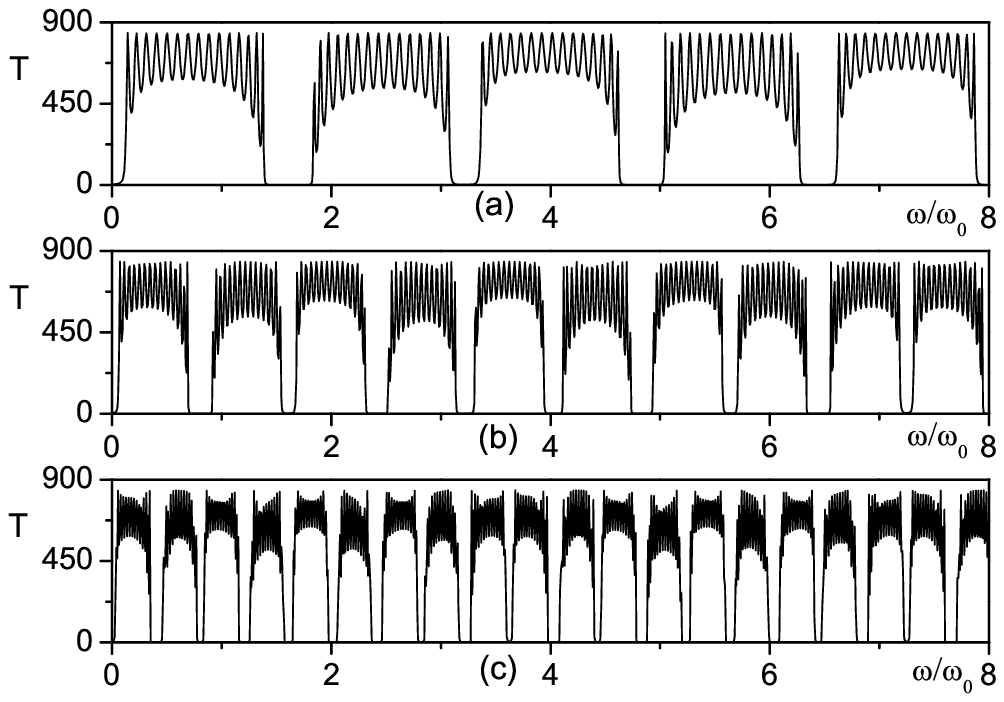}
\caption{Comparing the transmissivity of the GFPCs with different
optical thickness: (a)$\frac{\lambda_{0}}{8}$, (b)
$\frac{\lambda_{0}}{4}$ and (c)$\frac{\lambda_{0}}{2}$.}
\end{figure}

\begin{figure}[tbp]
\includegraphics[width=8.5cm]{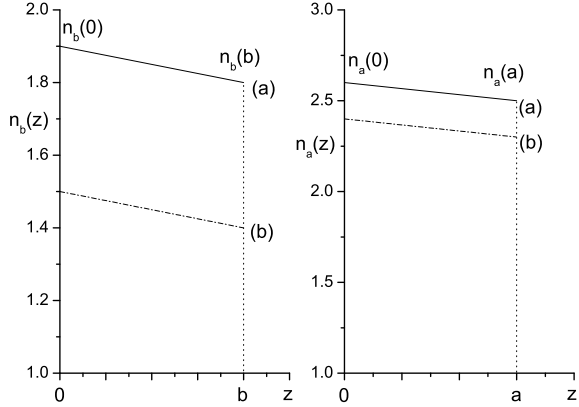}
\caption{The two groups line refractive index function in a
period: (a) the first group function and (b) the second group
function.}
\end{figure}

\begin{figure}[tbp]
\includegraphics[width=8.5cm]{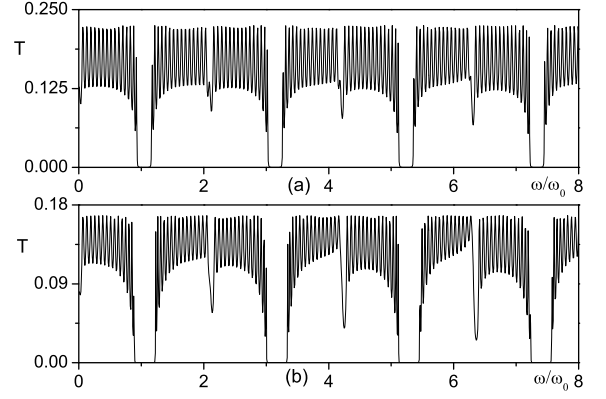}
\caption{Comparing the transmissivity of the GFPCs with different
line refractive index function: (a) the first group function and
(b) the second group function.}
\end{figure}

\begin{figure}[tbp]
\includegraphics[width=8.5cm]{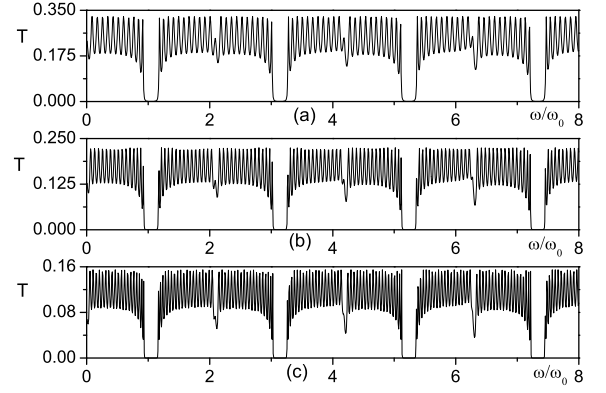}
\caption{Comparing the transmissivity of the function PCs with
different periodicity (a) N=12 (b) N=16 (c) N=20}
\end{figure}

\begin{figure}[tbp]
\includegraphics[width=8.5cm]{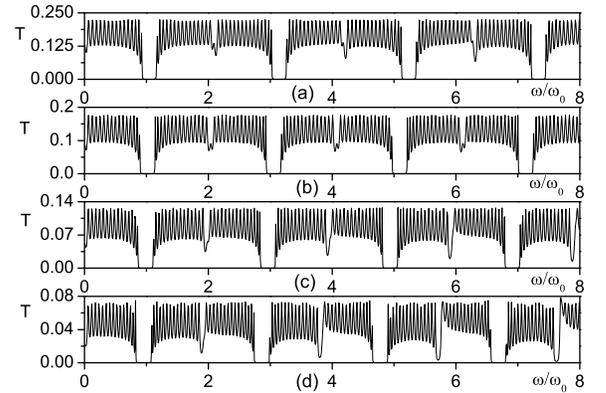}
\caption{Comparing the transmissivity of GFPCs with different
incidence angles: (a)$\theta_{i}^{0}=0$, (b)
$\theta_{i}^{0}=\frac{\pi}{6}$, (c)$\theta_{i}^{0}=\frac{\pi}{4}$
and (d) $\theta_{i}^{0}=\frac{\pi}{3}$.}
\end{figure}

{\bf 7. Conclusion} \vskip 5pt In summary, We have theoretically
investigated a new kind of general function photonic crystals
(GFPCs), which refractive index is a function of space position.
Based on Fermat principle, we achieve the motion equations of
light in one-dimensional general function photonic crystals, and
calculate its transfer matrix. We choose the linearity refractive
index function for two mediums $A$ and $B$, and find the
transmissivity of one-dimensional GFPCs can be much larger or
smaller than $1$ for different slope linearity refractive index
function. Otherwise, we study the effect of different incident
angles, the number of periods and optical thickness on the
transmissivity, and obtain some new results as follows: (1) when
$n_{b}(b)>n_{b}(0)$ and $n_{a}(a)>n_{a}(0)$, the peak value of
transmission intensity is much larger than $1$. (2) when
$n_{b}(b)<n_{b}(0)$, $n_{a}(a)<n_{a}(0)$, the transmission
intensity is much smaller than $1$. With the high and low
transmissivity of GFPCs, we can make amplification and decay
device of light. (3) For arbitrary medium $A$ or $B$, when its
slope of line refractive index function decrease the
transmissivity peak value decrease, the number of band gaps
decrease and the band gaps become narrow. (4) When the
transmissivity $T>1$, as the optical thickness increase, the
number of band gaps increase, and its width become narrow, while
the peak value of transmissivity is unchanged. (5) When the
transmissivity $T<1$, as the optical thickness increase, the
number of band gaps increase, and its width become narrow, while
the peak value of transmissivity is unchanged. (6) When the
transmissivity $T>1$, as the incident angles increase, the peak
value of transmissivity also increase, and the position of band
gaps red shift. (7) When the transmissivity $T<1$, as the incident
angles increase, the peak value of transmissivity decrease, and
the position of band gaps red shift. (8) When the transmissivity
$T>1$, as the number of period increase, the peak value of
transmissivity increase,while the position and width of band gaps
are unchanged. (9) When the transmissivity $T<1$, as the number of
period increase, the peak value of transmissivity decrease, while
the position and width of band gaps are unchanged. The new results
of GFPCs couldn't be found in conventional PCs, and we think the
GFPCs should be widely applied in the future.\\

\begin{figure}[tbp]
\includegraphics[width=8.5 cm]{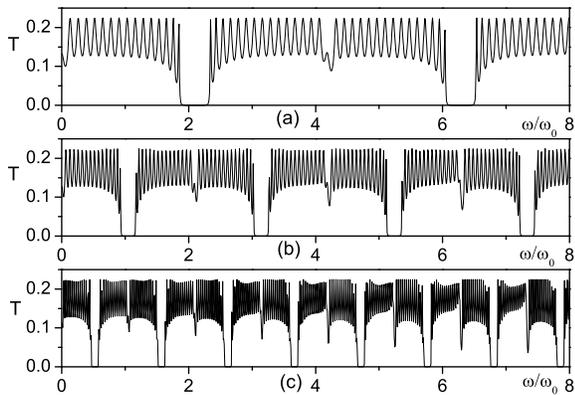}
\caption{Comparing the transmissivity of the GFPCs with different
optical thickness: (a)$\frac{\lambda_{0}}{8}$, (b)
$\frac{\lambda_{0}}{4}$ and (c)$\frac{\lambda_{0}}{2}$.}
\label{Fig1}
\end{figure}


\newpage
\begin{thebibliography}{10}

\bibitem{s1}
E. Yablonovitch, Phys. Rev. Lett. {\bf 58}, 2059 (1987).

\bibitem{s2}
F. Bordas, M. J. Steel, C. Seassal, A. Rahmani., Optics Express.
{\bf 15}, 10890 (2007).

\bibitem{s3}
P. Nedel, X. Letartre, C. Seassal, A. Auff¨¨ves, L. Ferrier, E.
Drouard, A. Rahmani, and P. Viktorovitch., Optics Express. {\bf
19} 5014 (2011).

\bibitem{s4}
C. Zinoni, B. Alloing, L. H. Li, F. Marsili, A. Fiore, L. Lunghi,
A. Gerardino, Yu. B. Vakhtomin, K. V. Smirnov, and G. N.
Gol'tsman., Appl. Phys. Lett. {\bf 91} 031106 (2007).

\bibitem{s5}
S. G. Johnson and J. D. Joannopoulos., Optics Express. {\bf 8} 173
(2001).

\bibitem{s6}
S. John, Phys. Rev. Lett. {\bf 58} 2486 (1987).

\bibitem{s7}
W. C. Stumpf, M. Fujita, M. Yamaguchi, T. Asano and S. Noda.,
Appl. Phys. Lett. {\bf 90} 231101 (2007).

\bibitem{s8}
V. S. C. Manga Rao and S. Hughes., Phys. Rev. Lett. {\bf 99}
193901 (2007).

\bibitem{s9}
G. Lecamp, P. Lalanne, and J. P. Hugonin., Phys. Rev. Lett. {\bf
99} 023902 (2007).

\bibitem{s10}
V. S. C. Manga Rao and S. Hughes., Phys. Rev B {\bf 75} 205437
(2007).

\bibitem{s11}
K. Inoue, M. Sasada, J. Kawamata, K. Sakoda, and J. W. Haus, Jpn
J. Appl. Phys. Part 2 {\bf 38} L157 (1999).

\bibitem{s12}
R. D. Meade, A. Devenyi, J. D. Joannopoulos, O. L. Alerhand, D. A.
Smith and K. Kash, J. Appl. Phys. {\bf 75} 4753 (1994).

\bibitem{s13}
T. Lund-Hansen, S. Stobbe, B. Julsgaard, H. Thyrrestrup, T.
S¨¹nner, M. Kamp, A. Forchel, and P. Lodahl., Phys. Rev. Lett.
{\bf 101} 113903 (2008).

\bibitem{s14}
S. J. Dewhurst, D. Granados, D. J. P. Ellis, A. J. Bennett, R. B.
Patel, I. Farrer, D. Anderson, G. A. C. Jones, D. A. Ritchie, and
A. J. Shields., Appl. Phys. Lett. {\bf 96} 031109 (2010).

\bibitem{s15}
K. Busch and S. John, Phys. Rev. Lett. {\bf 83}, 967 (1999).

\bibitem{s16}
A. Z. Sanchez and P. Halevi, J. Appl. Phys. {\bf 94}, 797 (2003).

\bibitem{s17}
E. Kuramochi, M. Notomi, S. Hughes, A. Shinya, T. Watanabe, and L.
Ramunno, Phys. Rev. {\bf B72}, 161318(R) (2005).

\bibitem{s18}
S. Hughes, L. Ramunno, Jeff F. Young , J. E. Sipe., Phys. Rev.
Lett. {\bf 94} 033903 (2005).

\bibitem{s19}
N. Le Thomas, H. Zhang, J. J¨¢gersk¨¢, V. Zabelin, and R. Houdr¨¦
, I. Sagnes and A. Talneau , Phys. Rev. {\bf B80} 125332 (2009).

\bibitem{s20}
M. Patterson, S. Hughes, S. Combri¨¦, N.-V.-Quynh Tran, A. De
Rossi, R. Gabet and Y. Jaouen., Phys. Rev. Lett {\bf 102} 253903
(2009).

\bibitem{s21}
J-K. Yang, H. Noh, M. J. Rooks, G. S. Solomon, F. Vollmer and H.
Cao., Appl. Phys. Lett. 98, 241107 (2011)

\bibitem{s22}
A. Figotin, A. Y. Godin and I. Vitebsky, Phys. Rev. B{\bf 61},
15523 (2000).

\bibitem{s23}
R. Martinez-Sala, J. Sancho, J. V. Sanchez, V. Gomez, J. Llinares
and F. Meseguer, nature {\bf 378}, 241 (1995).

\bibitem{s24}
J. V. Sanchez-Perez, D. Caballero, R. Martinez-Sala, C. Rubio, J.
Sanchez-Dehesa, F. Meseguer, J. Llinares and F. Galvez, Phys. Rev.
Lett. {\bf 80}, 5325 (1998).

\bibitem{s25}
J. D. Joannopoulus, S. G. Johnson, J. N. Winn and R. D. Meade,
Photonic Crystals. Molding the Flow of Light (Princeton University
press, Princeton, 2008).

\bibitem{s26}
D. Torrent, A. Hakansson, F. Cervera and J. Sanchez - Dehesa,
Phys. Rev. Lett. {\bf 96}, 204302 (2006).

\bibitem{s27}
D. Torrent, J. S¡äanchez-Dehesa, New. Jour. Phys. {\bf 9}, 323
(2007).

\bibitem{s28}
J. O. Vasseur, P. A. Deymier, B. Djafari-Rouhani, Y. Pen - nec and
A. C. Hladky-Hennion, Phys. Rev. B {\bf 77}, 085415 (2008).

\bibitem{s29}
Xiang-Yao Wu, Bai-Jun Bai, Jing-Hai Yang, Xiao-Jing Liu, Nuo Ba,
Yi-Heng Wu and Qing-Cai Wang, Physica E {\bf 43}, 1694 (2011).
\end{thebibliography}
\end{document}